\documentclass[aps,pre,preprint,showpacs,unsortedaddress,12pt]{revtex4-2}
\usepackage{amsmath}
\usepackage{graphicx}
\usepackage{epstopdf}
\usepackage{nicefrac}
\usepackage{color}

\usepackage{booktabs} 
\usepackage{colortbl} 
\usepackage{xcolor} 
\usepackage{xfrac}
\usepackage{makecell}
\usepackage[export]{adjustbox}
\usepackage{dblfloatfix}   
\begin{document}
		
		\author{Fatih Yasar}
		\email{fyasar@ou.edu} 
		\affiliation{Dept. of Chemistry \& Biochemistry, University of Oklahoma, Norman, OK 73019, USA}
		
		\author{Alan J. Ray}
		\email{alanjray@ou.edu} 
		\affiliation{Dept. of Chemistry \& Biochemistry, University of Oklahoma, Norman, OK 73019, USA}
		
		\author{Ulrich H. E. Hansmann}
		\email{uhansmann@ou.edu} 
		\affiliation{Dept. of Chemistry \& Biochemistry, University of Oklahoma, Norman, OK 73019, USA}
		
		\title{Resolution Exchange with Tunneling\\ for Enhanced  Sampling of Protein Landscapes} 
		\date{\today}
		
		\begin{abstract}
			Simulations of protein folding and protein association  happen 
			on timescales that are orders of magnitude  larger than what can typically be covered in all-atom molecular 
			dynamics simulations. Use of low-resolution models alleviates this problem but  may reduce the accuracy 
			of the simulations. We introduce a replica-exchange-based  multiscale sampling technique that 
			 combines the faster sampling in coarse-grained simulations with the potentially higher accuracy 
			of all-atom simulations. After testing the efficiency of our Resolution Exchange with Tunneling (ResET) 
			 in simulations of the Trp-cage protein, an often used model to evaluate sampling techniques in protein simulations,  
			 we use  our approach to compare  the landscape of wild type and A2T mutant  A$\beta_{1-42}$ peptides.
			 Our results suggest a mechanism by that the mutation of a small hydrophobic Alanine (A)  into a bulky polar
			  Threonine (T) may interfere with the self-assembly of A$\beta$-fibrils. 
		\end{abstract} 
		
		\maketitle

		\section{Introduction}
		While molecular dynamics is  now commonly used to study folding, association and aggregation
		of proteins and other biological macromolecules   
		\cite{dill_2012,dill_2008,dobson_2003,han_2010, chiti,Stroud7717,hansmann_2013_1,Todd_2011,Eisenberg},
		biochemical processes such as the formation of amyloid fibers from monomers \cite{Eisenberg,chiti,seeding_eisele}
		often  occur on timescales \cite{Weise_2010,seeding_eisele} 
		that exceeds what can be covered in all-atom  simulations.  Coarse-graining, i.e., lowering the 
		resolution of a system  \cite{cg_review,han_2010,pace_2007,rosetta,CABS,UNRES},
		allows one to reduce the computational difficulties and    to access  timescales not obtainable
		 to the  fine-grained all-atom models \cite{han_2010,cg_review},  but it often results in lower accuracy. 
		 This is because the  smaller number of degrees of freedom  lowers  the  entropy
		 of the system, and it is difficult to compensate for this reduction   by modifying   the enthalpic contributions accordingly \cite{cg_review}. 
		Multiscale  techniques  try to combine the advantages of fine-grained models (that are more accurate but costly to evaluate) 
		with that of   coarse-grained models  (which are less detailed but enable  larger time steps).  
		
		One example is   
		Resolution Exchange \cite{Lyman_2006_re} where the   replica-exchange protocol \cite{H97f} is used
		to induce a walk in resolution space. In the same way that for  Replica-Exchange molecular dynamics (REMD) \cite{H97f,Sugita_Okamoto} 
		the walk in temperature space leads   to  faster sampling at low temperatures, 
		enables  exploration  of resolution space  a faster convergence of simulations at  an all-atom level \cite{Lyman_2006_re,Lyman_2006}. 
		However, the replica-exchange step requires reconstruction of the fine-grained degrees of freedom 
		of a previously coarse-grained configuration,  for instance, by adding side chains   to a conformation 
		that  was described prior only  by the backbone. Various 
		approaches  \cite{Liu_2008,Liu_2007,lwin_2005,Lyman_2006_re,Lyman_2006}  have been developed to address 
		this problem, but often they result in high energies of the proposal configuration (and therefore low acceptance rates)  
	   	\cite{Lyman_2006,Liu_2008}   or introduce biases \cite{Liu_2007,lwin_2005}.
		
		The dilemma can be alleviated by introducing   a potential energy made of three terms:
		\begin{equation}
			E_{pot} = E_{FG} + E_{CG} + \lambda E_{\lambda}~.
			\label{3energy}
		\end{equation}
		The first term is the energy  $E_{FG}$  of the protein system and the surrounding environment as described 
		by an \textit{all-atom} (\textit{fine-grained}) model.
		The second term $E_{CG}$ describes the same  system by a suitable \textit{coarse-grained} model.  
		Both models are coupled by a {system-specific  penalty} term $E_{\lambda}$~\cite{Moriatsu,Chen} that measures the similarity
		between  the configurations at both levels of resolution, with the strength of coupling controlled 
		 by a replica-specific  parameter $\lambda$.   Hence, 
		Hamilton Replica Exchange \cite{Takada_2002, hansmann_2005} of the above defined multiscale system leads 
		to an  exchange of information  between fine-grained and coarse-grained models, with measurements taken  
		at the replica where $\lambda=0$.	However, while avoiding the problem of steric clashes  in resolution exchange,
		 the exchange probability  is often still  small \cite{hansmann_2007}, and  	 the resulting need  to use multiple replica
		  to bridge the two levels of resolution makes this approach not appealing.  
		  
		   As an alternative,  we propose here a
		Resolution Exchange with Tunneling (ResET) approach that requires only two replicas. Working and efficiency of our approach
		 is tested in simulations of   the Trp-cage \cite{Neidigh:2002aa,roitberg_2002} miniprotein (Protein Data Bank (PDB)  Identifier: 1L2Y),
		an often used model for testing new sampling techniques.   As a first application we use in the second part ResET  to compare the landscape 
		of A$\beta_{1-42}$ wild type peptides, implicated in Alzheimer's disease,  with that of  A2T mutants which seems to protect
		against Alzheimer's disease~\cite{hardy_2002,maho_2002, selkoe_2002}. Our results suggest a mechanism by that the mutation 
		of a small hydrophobic Alanine (A) into a bulky polar
			  Threonine (T) may interfere with the self-assembly of A$\beta$-fibrils, decreasing  the chance for formation of  the A$\beta$-amyloids
			  that are a hallmark of Alzheimer's disease ~\cite{A2T,maloney_2014, benilova_2014}.

		\section{Resolution Exchange with Tunneling}

		Resolution Exchange with Tunneling (ResET) utilizes 	two replica, each containing both a
		coarse-grained and a fine-grained representation of the system. On each replica, both representations  
		evolve separately by molecular dynamics. 
		On the first replica, A,  is  the \textit{coarse-grained model} in a configuration $A_{CG}$ and has a potential  energy $E^{pot}_{CG}(A_{CG})$
		and a kinetic energy $E^{kin}_{CG} (A_{CG})$.  On the other hand, the  fine-grained model is in a configuration $A_{FG}$  
		that has a kinetic energy $E^{kin}_{FG} (A_{FG})$  and a potential energy  $E^{biased}_{FG}(A_{FG})$ which depends on the
		configuration $A_{CG}$ of the coarse-grained model by  
		$E^{biased}_{FG}(A_{FG}) = E^{pot} _{FG}(A_{FG}) + \lambda_1 E_{\lambda}(A_{CG},A_{FG})$. Hence, the  
		two models on this replica  interact only  by the  term $\lambda_1 E_{\lambda}(A_{CG},A_{FG})$  that biases the fine-grained model, 
		but are otherwise invisible to each other. The effect of this biasing term is that  configurations of the fine-grained model are favored 
		which resemble the coarse-grained model configuration,  with the strength of the bias controlled by parameter  $\lambda_1 $. 
		The opposite situation is found on the replica B.
		Here lives an independent fine-grained model with configuration $B_{FG}$ that has a   potential energy $E^{pot}_{FG}(B_{FG})$ 
		and kinetic energy $E^{kin}_{FG} (B_{FG})$, while, on the other hand, the configuration  $B_{CG}$ of the coarse-grained model 
		  has a kinetic energy $E^{kin}_{CG} (B_{CG})$ and a potential energy 
		$E^{biased}_{CG}(B_{FG},B_{CG}) = E^{pot} _{CG}(B_{BG}) + \lambda_2 E_{\lambda}(B_{CG},B_{FG})$  
		that depends  on the fine-grained model by a term $\lambda_1 E_{\lambda}(A_{CG},A_{FG})$. This biasing term now   
		ensures that on replica B  the coarse-grained configuration resembles the one of the fine-grained model.

		 While the time step for integrating fine-grained and coarse-grained models may differ, they have  to be the same 
		 for the corresponding models on both replicas.	This is because 
		 after a certain number of molecular dynamics steps  a decision is made on whether to replace on the  replica  B
		 the configuration $B_{FG}$ in the unbiased fine-grained model  by the configuration $A_{FG}$ of the auxiliary (biased)
		  fine-grained model of the  replica A.  This replacement goes together with a re-weighting of the velocities $v_{FG} (A_{FG})$
		  such that $\hat E^{kin}_{FG} (A_{FG}) = E^{kin}_{FG}(B_{FG})$, and  is accepted with probability:
\begin{equation}
	w (B \rightarrow A)  = \min\left(1,\exp(-\beta  (E^{pot}_{FG}(A_{FG}) - E^{pot}_{FG}(B_{BG}) 
	-  \lambda_1 E_{\lambda} (A_{FG},A_{CG}) -\Delta E^{kin}_{FG} )\right)
	\label{ResET1}
\end{equation}	
with $\Delta E^{kin}_{FG} = E^{kin}_{FG}(A_{FG}) - E^{kin}_{FG}(B_{FG})$. The re-weighting of the velocities 
 and the Metropolis-Hastings acceptance 
criterium accounts for the fact  that the  proposal configurations $A_{FG}$  are generated on replica A by a  biased process, i.e., it corrects for the
resulting  skewed  probability with  which the  configuration  $A_{FG}$ is proposed as a replacement for $B_{FG}$. 

At other times, the the coarse-grained configuration $A_{CG}$ 
on replica A is replaced by the configuration $B_{CG}$
of the biased coarse-grained model on replica B with probability:
\begin{equation}
	w(A \rightarrow B) = \min \left(1,\exp(-\beta  (E^{pot}_{CG}(B_{CG}) - E^{pot}_{CG}(A_{BG}) 
	- \lambda_2 E_{\lambda} (B_{FG},B_{CG})  -  \Delta E^{kin}_{CG})\right) 
\label{ResET2}
\end{equation}
with $\Delta E^{kin}_{CG} = E^{kin} _{CG} (B_{CG}) + E^{kin}_{CG} (A_{CG})$. Re-weighting the velocities of configuration $B_{CG}$ such that 
$\hat E^{kin}_{CG} (B_{CG}) = E^{kin}_{CG}(A_{CG})$,  and the Metropolis-Hastings acceptance 
criterium are again to correct for the skewed probability by which the  configuration  $B_{CG}$ is proposed. 
	
Note that the update of the unbiased coarse-grained configuration on replica A also changes the $E_{\lambda}$ biasing term 
in the ancillary fine-grained configuration, as does the update of the unbiased fine-grained configuration on replica B changes 
the corresponding biasing  term in the steered coarse-grained configuration. In order to minimize this disturbance, we also 
rescale the velocities in the biased models such that the change in kinetic energy compensates for the change in $E_{\lambda}$.

We remark that in software packages such as GROMACS~\cite{gromacs} it is sometimes simpler to separate the biased and unbiased models onto 
different replicas. In this case one would have four replicas, with a possible  distribution of the models sketched in the table below.
\begin{center}
	\resizebox{\textwidth}{!}{%
\begin{tabular}{lccccc}
	Model                                        &replica&Potential Energy& Kinetic Energy & Lambda         & Lambda Energy\\
	\hline
	unbiased fine-grained model      & 0       & $P_0$               & $K_0$              &                       &                                   \\
	biased fine-grained model          & 1       & $P_1$               & $K_1$              & $\lambda_1$ & $E_{\lambda} (1,3)$  \\
	biased coarse-grained model      & 2       & $P_2$               & $K_2$              & $\lambda_2$ & $E_{\lambda} (0,2)$  \\
	unbiased coarse-grained model  & 3       & $P_3$               & $K_3$              &                       &                                  \\ 
	\hline
\end{tabular}}
\end{center}
In this implementation, the replica 0 and 2, and replica 1 and 3, communicated   during the molecular dynamics evolution of the configurations;
and the ResET move  replaces the configuration of  replica 0 by that of replica 1, and/or the configuration on replica 3 by that of replica 2. 

\section{Materials and Metods}		
 \subsection{Set up of the ResET simulations}

Our simulations utilize a modified version of the GROMACS \cite{gromacs} molecular package available from the authors.
Initial tests of the working and efficiency are for the Trp-cage protein \cite{Neidigh:2002aa,roitberg_2002}, an often used system
 for evaluating new algorithms. In order to compare our simulations with previous studies,  we follow closely the set-up 
 of  Han et al \cite{han_2013}  for the coarse-grained model,  and that of  Kouza et al \cite{hansmann_2011}  for the fine-grained  model.
 Hence,  our coarse-grained Trp-cage protein model is described  by PACE force-field \cite{han_2010}, with  the uncapped protein
 solvated by  1118 MARTINI \cite{marrink_2007} coarse-grained water molecules,  and buffered 0.15M $Na^+$ and $Cl^-$ ions, 
 in a cubic box  of length 5.18 nm, leading to a total of 1313 coarse-grained  particles.   On the other hand, in our fine grained model 
 is the N-terminus capped by an acetyl group and at  C-terminus by methylamine, leading to a total number of 313 atoms for the protein
  that are solvated with 2645 extended  simple point charge (SPC/E \cite{Berendsen_1987}) water molecules in a cubic box with an edge 
  length of 4.4 nm.  One chlorine ion ($Cl^-$) is  added to neutralize the system. Hence, the system contains 8249  fine-grained particles,
 with the interactions between them described by  the AMBER94 force-field \cite{kollman_1995}. 		

As a first application we compare in the second part of this study the ensemble of configurations sampled by ResET simulation 
of  A$\beta_{1-42}$ wild type and A2T mutant  peptides.  While aggregates of the wild type A$\beta$-peptides  are implicated
 in Alzheimer's disease, the A2T mutant appears to be protective, i.e, reducing the probability for acquiring the disease.  We use
 in our simulations for both wild type and mutant as coarse-grained  model   the  MARTINI force-field \cite{marrink_2007},
which   is  computationally efficient and has been already used earlier in A$\beta$ simulations   \cite{zheng_2015, poma_2019}.
Here, the main chain of each amino acid is represented by one bead, and the side chains by up to four beads 
depending on the size of the amino acid. Our wild type protein thus contains  91 beads, and the mutant  92 beads. 
Each peptide is placed in a cubic box and solvated with the MARTINI-CG water molecules  represented by single beads. 
Together with 3 $Na^+$ MARTINI-ion beads and a box size of 7.16 nm (wild type) and 7.24 nm A2T mutant) 
we arrive at 2925  and 3189 particles, respectively.	On the other hand, the fine-grained representations of wild type and mutant peptides  are
modeled by the CHARMM36 force-field \cite{charmm36} which we found in previous work to be 
		efficient for simulations of intrinsically disordered and amyloid-forming proteins. The N- and C-termini are capped with Methyl groups. 
		The protein is placed in the center of a cubic box using a 1 nm distance between the atoms of the protein and box.  
		The each system is   solvated with  TIP3 water molecules \cite{jorgensen_1983} and neutralized with 3 $Na^+$ ions.  
		This leads to a box size of 7.5 nm and a total number of 41412 particles for the wild type. Correspondingly, 
		we get a box size of 7.6nm and a total number of 44509 particles for the mutant.
		
In simulations of both the Trp-cage protein and the A$\beta$-peptides we use for both fine-grained and coarse-grained models 
	shift functions with a cut-off of 1.2 nm  in the calculations of Coulomb and van der Waals  interactions.		
	 Because of periodic boundary conditions we  employe    Particle mesh Ewald (PME) \cite{pedersen_1995}  summation 
	 to account for long-range electrostatic interactions. Hydrogen atoms and bond distances are constraint in the fine-grained model 
	 by the LINCS algorithm \cite{LINCS}.  		
	Equations of motion  are  integrated using a leap-frog algorithm, with a time step of 2 fs for both the fine-grained model and coarse-grained model. 
	The v-rescale thermostat ~\cite{Bussi}  with a coupling time of 0.01 ps is used to maintain  the temperature in the coarse-grained models, while a
	Nose-Hoover \cite{Nose, Hoover} thermostat with the coupling time of 0.5 ps controls the temperature in the fine-grained models. 		 
	  				
A key element of the ResET sampling technique is the
restraining potential $E_{\lambda}$ which quantifies the similarity between fine-grained and coarse-grained configurations. 
		In our case, we choose a function of the form \cite{Chen}.
		\begin{equation} 
			E_{\lambda}(q_{FG},q_{CG}) = 
			\left\{ \begin{array}{ll}
				{\displaystyle \frac{1}{2}\left(\Delta^2 \left(i,j\right)\right)          }                               \quad  & -ds  < \Delta \left(i,j\right) <  ds    \\
				{\displaystyle   A + \frac{B}{\Delta^S \left(i,j\right)} + f_{max} \Delta (i,j)  }   \quad   &  \Delta (i,j) > ds     \\
				{\displaystyle   A +  \frac{B}{\Delta^S (i,j)} \left( -1 \right)^{S} -  f_{max} \Delta (i,j)  }  \quad  &  \Delta  (i,j) < - ds             
			\end{array} 
			\right.
			\label{lambdaenergy}
		\end{equation}
		\noindent
		where $q_{FG}$ are the coordinates of atoms in the fine-grained model and $q_{CG}$ the ones 
		in the coarse-grained model.  $\Delta(ij) = \delta_{FG}(ij) - \delta_{CG}(ij)$ is the difference between  the distances ($\delta(ij)$) 
		measured  in either the fine-grained or the coarse-grained  models between the C$_{\alpha}$-atoms  i and j.
		The control parameter $f_{max}$  sets the maximum force as $\Delta \left(i,j\right) \rightarrow \infty$ and 
		$S$ determines how fast this value is realized. 
		The parameters $A$ and $B$ are included to ensure continuity of $E_{\alpha}(q_{fg},q_{cg})$ and it's first derivative 
		at  values  where $\Delta \left(i,j\right) = \pm ds$,
		i.e., where the functional form of \ref{lambdaenergy} changes. These parameters are thus computed by 
		\begin{equation}
			A = \left( \frac{1}{2} + \frac{1}{S}  \right)ds^2    -   \left( \frac{1}{S} 
			+ 1   \right)f_{max}ds  \quad and \quad B = \left(\frac{ f_{max}  - ds }{S} \right)ds^{S+1}.
		\end{equation}
        In the ResET simulations is the biased  fine-grained model on replica A coupled to the unbiased  coarse-grained model by a parameter
        $\lambda_1= 0.5$, while  on replica B the biased coarse-grained models is coupled to the free fine-grained models by a parameter 
        $\lambda_2 = 2.5$.  The ResET replacement move is tried every 250 ps,
        with the bias-correction factor $ \lambda_1 E_{\lambda} (A_{FG},A_{CG}) -\Delta E^{kin}_{FG}$ limited to the interval (0,100), and
        on replica B $ \lambda_2 E_{\lambda} (B_{FG},B_{CG})  -  \Delta E^{kin}_{CG})$ to the interval  (0,20), choices that we found in preliminary 
        test runs leading to increased numerical stability.

  		\begin{table}[]
  			\caption{Simulation details}
  			\begin{center}
  			\resizebox{\textwidth}{!}{%
  			\begin{tabular}{lccccccc}
  				\hline
  				& \multicolumn{1}{c}{} & Trp-cage &                 &  &             & A$\beta_{1-42}$                    &                 \\ \cline{2-4} \cline{6-8} 
  				Method            & Force-Field  &Sampling No&Time (ns)      &  & Force-Field          &Sampling No&Time(ns) \\ \hline
  				Canonical FG      & AMBER94      &3         &5000             &  & CHARMM36             &     $---$    &   $---$             \\
  				REMD      FG      & AMBER94      &1         &200             &  & $---$                & $---$     &   $---$            \\
  				ResET     FG+CG~~~&AMBER94+PACE~~&6         &200(1000)        &  & CHARMM36+MARTINI v2.2&     2     &    100(500)     \\ \hline            
  			\end{tabular}}
  			\label{sim_details}
  			\end{center}
  		\end{table}
  
	Start structures for both fine-grained and coarse-grained models are generated by heating up the  experimental structures 
	of PDB-ID: 1L2Y (Trp-cage) and  PDB-ID: 1Z0Q (A$\beta_{1-42}$) \cite{tomaselli_2006, A2T} to 500 or 1000 K 
	in short molecular dynamics simulations under NVT conditions (0.5 ns and 1 ns),
	and cooling them down to the respective temperatures (with the exception of the REMD simulations is this 310 K). Simulations of the various systems
	start from the so-generated configurations and are performed in the NVT ensemble, with the simulation details listed in Table \ref{sim_details}.

  For most of our analysis we use GROMACS tools \cite{gromacs}  such as gmx-rms which calculates the root-mean-square deviation (RMSD) 
  and the root-mean-square fluctuations (RMSF) of residues  with respect to an initial configuration. For  visualization we use the VMD software \cite{vmd}, which we also use to calculate the solvent accessible surface area (SASA) using a probe radius of 1.4 \AA.	
  Other quantities are calculated with in-house programs and defined in the manuscript. 
   An example are dynamic cross-correlation  
  maps which are  calculated using  the definition of \cite{swaminathan_1991, hansmann_2021}:
	   \begin{equation}
	  		C(i,j) = \frac{\langle\Delta {\bf r}_{i} .\Delta  {\bf r}_{j}\rangle}{\langle\Delta {\bf r}_{i}^2\rangle  \langle\Delta {\bf r}_{j}^2\rangle} . 
  \end{equation}	  
	  where $ \Delta {\bf r}_{i}$    and $\Delta {\bf r}_{i}$ are  the displacement vectors of $i$-th  and  $j$-th residues of the system 
	  and angle brackets represent ensemble  averages. Positive values mark correlated motions of the respective residues
	   while negative values indicate anti-correlated motion.

  		\section{Results and Discussion}
  		\subsection{Efficiency of ResET}
\begin{figure}[h!]
	\centering
	\includegraphics[width=0.9\textwidth]{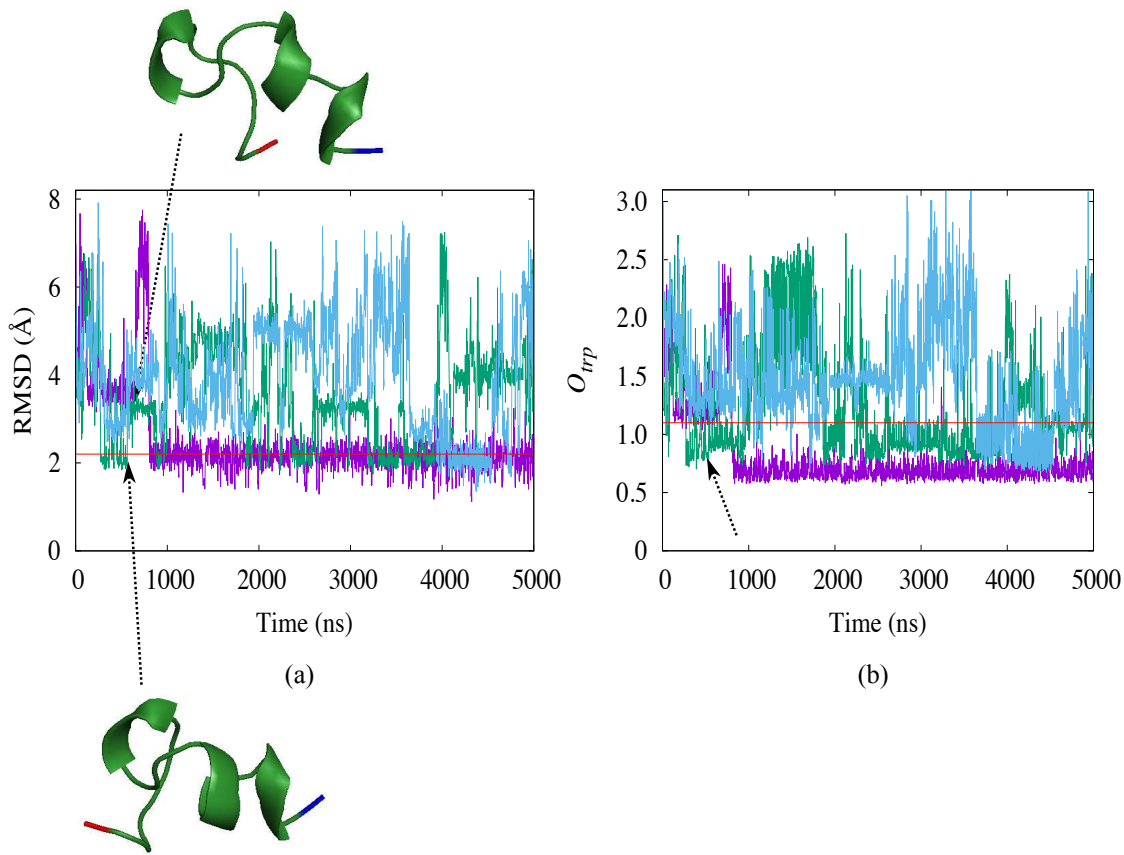}
     \caption{The time evolution of RMSD (a)   and   folding parameter $O_{trp}$ (b) as measured in regular molecular dynamics simulations  
     at T=310 K. Trajectory 1 is drawn in purple, trajectory 2 in green   and trajectory 3 in blue. The two snapshots are taken from   trajectory 2
      at 601.0 ns (snapshot at the bottom) and 602.3 ns (snapshot on top). Both
     snapshots show similar configurations while  the RMSD changes from  2.0 \AA ~ to 3.6 \AA .   N- and C-terminal residues
      in the snapshots are marked in  blue and red color, respectively.}
     \label{Fig-CANON-RMSD}
\end{figure}

In order to test the working and efficiency of our multiscale approach ResET, we perform first   simulations of   
the Trp-cage \cite{Neidigh:2002aa,roitberg_2002} miniprotein,	an often used model for testing sampling techniques. 
 Choice of this system, with which we are familiar from previous work,  therefore allows  a  direct comparison 
		with past simulations. An example are the replica  exchange molecular dynamics (REMD) 
		simulations of Ref.~\cite{garcia_2007, hansmann_2011}, where   40 replicas of equal volume are simulated at 40 temperatures
		spanning a range from T= 280 K to T=540 K. Configurations are exchanged between neighboring temperatures according
		to a generalized Metropolis criterium, leading to a random walk in temperature  that allows replicas to find local minima 
		(when at low temperatures) and escape out of them (when at high temperatures). The net-effect is an enhanced sampling
		at the target temperature. Defining a configuration as native-like if the   root-mean-square deviation (RMSD) to the PDB-structure 
		(PDB-ID: 1L2Y) of less than 2.5 \AA,  we find at T=310 K native-like configuration with a frequency of 87 \%, using the more
		restrictive criterium of a RMSD smaller than 2.2 \AA, the frequency reduces to 55\%. Note, that these frequencies do
		not change beyond statistical fluctuations once the REMD simulation has reached 50 ns, and we therefore neglect 
		the first 50 ns of our  200 ns long trajectories when calculating the frequencies. While these frequencies are similar 
		to the ones  observed in 	earlier work \cite{garcia_2007, hansmann_2011}, we suspect that our values overestimate the
		 frequency of folded configurations that reside at a certain time at T=310 K. This is because 
		 		 the systems are simulated at each temperature with the same  volume. This volume, while sufficiently large  at
		the target temperature may at  the higher temperature  
		suppress extended configurations, therefore artificially stabilizing folded configurations.	For this reason, we prefer 
		to compare our ResET simulations instead directly with  	
		regular constant temperature molecular dynamics, simulating the Trp-cage protein in three  independent 
		trajectories at T=310 K over 5000 ns, a value that is comparable to the experimental measured folding times
		 of around   4$\mu$ ~\cite{Oui_fold}.
               The RMSD as function of time  is shown  	for  all three trajectories in Figure \ref{Fig-CANON-RMSD}a.

		Visual inspection of the three trajectories points to another problem. For a small protein such as Trp-cage is  the RMSD not
		 good measure for similarity as configurations that appear as similar by visual inspection may differ by relatively large RMSD values. 
		 This can be seen, for instance, in the second trajectory where at  around 600 ns the RMSD increases from  2.0 \AA\  to 3.6 \AA, 
		 i.e., from native-like to configurations to one considered no longer native-like according to the above definition 
		 of a native configuration (i.e., having a RMSD of less   than 2.5 \AA).   However, visual inspection shows
		 that the  molecule keeps its native-like fold, see the  corresponding configurations also shown in the Figure.  
		This contradiction between our RMSD-based definition and visual inspection  made us configurations while the RMSD consider another quantity as measure for similarity.
		The two main characteristics of the Trp-cage native structure are its two helices (residues  2-9 and 11-14),
		 and the contact between residues  6W (a Tryptophan)   and residue 18P (a Proline). 
		 Hence we define as marker for Trp-cage folding a new quantity: 
		\begin{equation}
		O_{trp} = d_{6-18} + 1/(n_H+1)
		\end{equation}
		Here, $d_{6-18}$ is the difference between residues 6W and 18P, and $n_H$ the number of residues that have
		dihedral angles as seen in a helix.
		The time evolution of this quantity in Figure \ref{Fig-CANON-RMSD}b		shows that the
		new coordinate allows indeed a better  description of the folding transitions, as its behavior differs less from the
		visual inspection. Especially, we do not see  for the second trajectory  at 600 ns  the false signal for non-native configurations 
		that we see in the RMSD plot.
		 Comparing  $O_{trp}$ as function of time with visual inspection 
		 of configurations along the trajectories suggests that folded configurations are characterized by values of $O_{trp} < 1$, 
		 and we use in the following this definition to quantify frequencies of folded configurations. 
		 
		 With this definition, we observe 
		the first folding event  at t=11.6 ns (in trajectory  2), and the systems stays folded for about 600 ns before unfolding again.
		For trajectory  1 folding is observed at t=800 ns, and no folding is observed within 3500 ns in the third trajectory where the protein
		 unfolds afterwards  again at about 4500 ns.  As a consequence, we find between 250 ns and 500 ns folded configurations with a frequency 
		 of about 26\% 
		 and between  750 ns and 1000 ns, with about 49\%. 
		 The frequencies increase only slowly as the simulations proceed, and between 3000 ns and 5000 ns we find native-like
		 configurations with about 58\%. 
		 The above numbers are consistent with the experimentally measured folding times  of about 4$\mu$ ~\cite{Oui_fold}.

\begin{figure}[h!]
	\centering
	\includegraphics[width=1.0\textwidth]{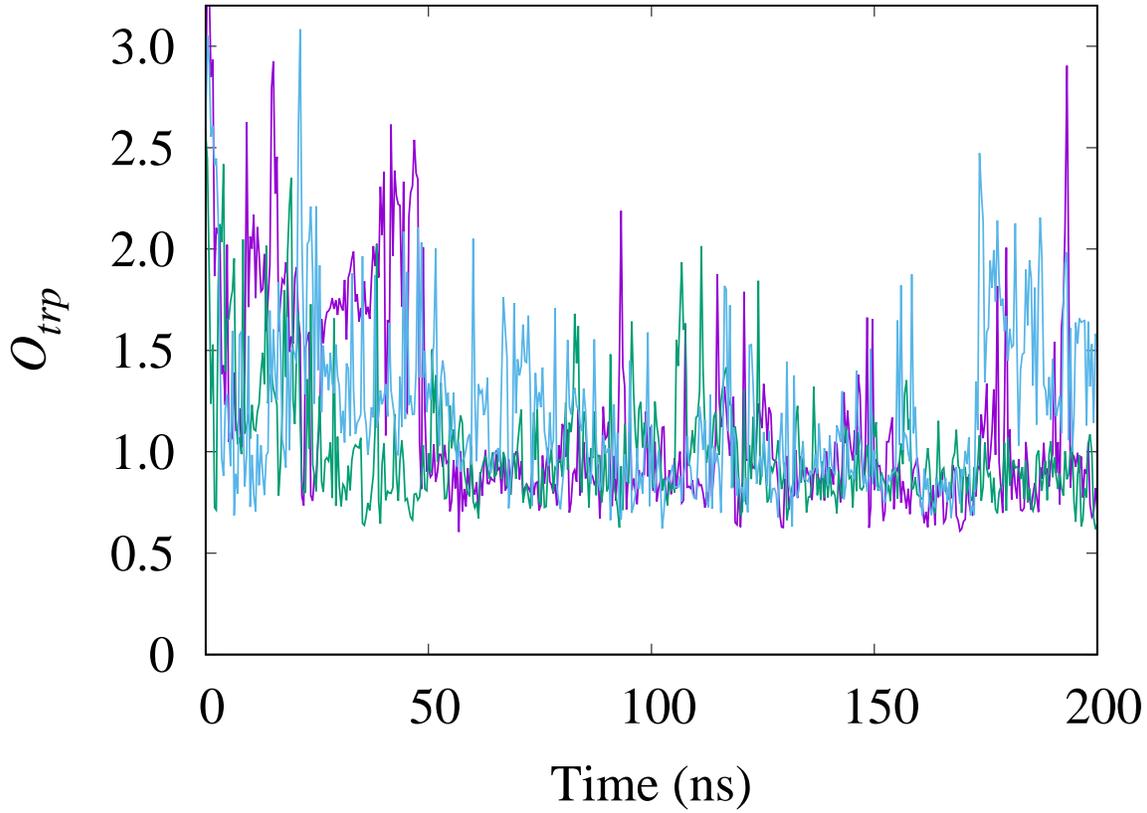}
	\caption{The time evolution of  the order parameter  for 100-20 kJ/mol. Trajectory 1 is drawn in purple, trajectory 2 in green   and trajectory 3 in blue.}
	\label{Fig-RESET-ORDER}
\end{figure}

		How does our new multiscale method fits in this discussion?
		The time evolution of our marker function $O_{trp}(t)$ is shown in 
		Figure \ref{Fig-RESET-ORDER}. Native-like configurations according to our criterium are observed after around 30 ns, 
		and between 50 ns and 100 ns 
		seen with a frequency of about 59\%. 
		The frequencies do not change much as the simulation progresses, and between 150 ns and 200 ns 
		are native-like configurations  observed with 65\%. 
		We remark that these frequencies do not depend  on the choice of parameters with which we scale the 
		$\lambda$ energy contribution in the ResET update.
		
		These frequencies for folded configurations are similar to what 
		is seen in long-time canonical runs,  but require shorter simulation times. Hence, our simulations of the Trp-cage protein
		indicate that our new multiscale simulation method leads indeed to an increase in sampling efficiency. If we take as criterium for
		the comparison the time it takes to have (on average) about 50\% of configuration folded (about 800 ns  for the canonical runs 
		and 50 ns for the ResET run) we find that ResET is about 16 times faster than the canonical simulations. 
		While the gain in efficiency will depend on the 
		specifics of the coarse-grained model (i.e. how much faster it samples the configuration space) and its coupling to the 
		physical force-field,  our data demonstrate the faster sampling properties of our multiscale approach.

		\subsection{Comparing A$\beta$ wild type and A2T mutant}
		
		Our evaluation of the sampling efficiency of ResET relies  on a rather simple test case. As a more interesting first application, we
		use in the second part  our sampling technique to compare the ensembles of wild type and A2T mutant  A$\beta_{1-42}$ peptides.
		Fibrils containing A$\beta_{1-40}$ or the more toxic A$\beta_{1-42}$ are a hallmark of Alzheimer's disease and the focus 
		of intense research \cite{hardy_2002}.  A large number of familial mutations are known that worsens the symptoms of Alzheimer's 
		disease or hasten its outbreak  \cite{maho_2002, selkoe_2002}, but there have been also  mutations  identified that are protective,
		i.e. lower the risk to fall ill with Alzheimer's disease.  One example is the mutant A2T 
		where the second residue (counted from the N-terminus) is changed from a small hydrophobic Alanine (A) into a bulky polar  
		Threonine (T) \cite{A2T}.  It has been not yet established  why this mutation is protective \cite{maloney_2014, benilova_2014}, 
		but one possibility is that this mutation
		alters the pathway for amyloid formation, for instance, by making it more difficult to form aggregates. In order to test this hypothesis 
		we simulate 	here  A$\beta_{1-42}$ wild type and A2T mutant monomers, and compare the ensembles of sampled configurations
		for their aggregation propensities.	
		
\begin{figure}[t!]
	\centering
	\includegraphics[width=1.00\textwidth]{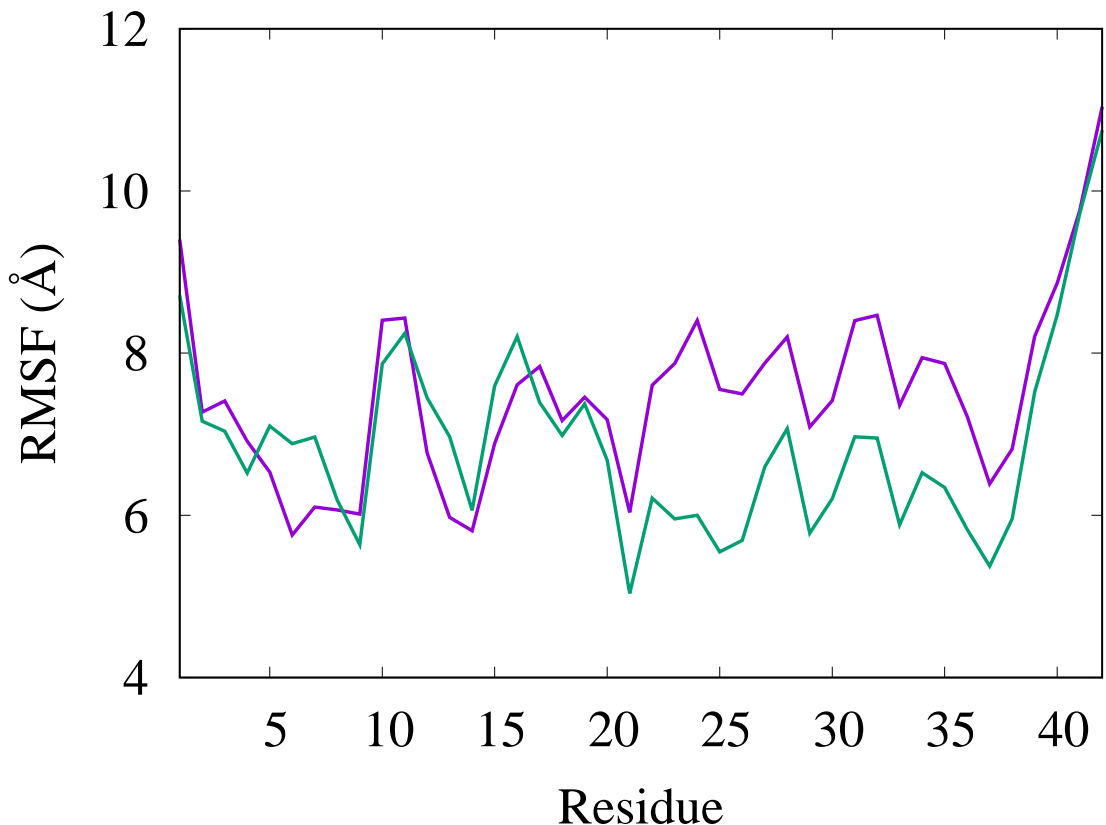}
	\caption{ Root-mean-square-fluctuations of residues  in either wild type (purple) or A2T mutant  (green)  A$\beta_{1-42}$ peptides. 
	               Only heavy atoms are considered in the calculation, and the first 50 ns of the 100 ns trajectories discarded to
	               allow for convergence of the simulations.}
	\label{Fig:rmsf}
\end{figure}

		Under physiological conditions are A$\beta$-peptides intrinsically disordered,  and we 
		do not expect the appearance of folded structures. Instead, we assume that the ensemble of configurations contains 
		such with transiently formed $\beta$-strands that would encourage aggregation. We conjecture  that such transient ordering appears more 
		often for  wild type A$\beta_{1-42}$ than for the A2T mutant peptides. In order to identify these differences in local ordering, we have 
		measured  the root-mean-square-fluctuations (RMSF) of residues for both cases, taking as reference structure the corresponding 
		start configuration, but discarding for the calculation of the RMSF the first 50 ns of the simulation. The RMSF is chosen because 
		this quantity describes the flexibility of residues
		or segments of the protein, and the more flexible a segment is the less likely will it be involved in forming stable structures.
		 Our data are shown in Figure \ref{Fig:rmsf}, and while there are only small differences for the first 20 residues between wild type
		 and mutant, the situation is different for the C-terminal half of the chain. For residues 21-37 is the RMSF considerably lower for
		 the mutant than for the wild type. We remark that this picture does not change if we recalculate the RMSF, including now all heavy
		 atoms (i.e, not only backbone but also side-chain atoms).

\begin{figure}[t!]
		\centering
		\includegraphics[width=1.0\textwidth]{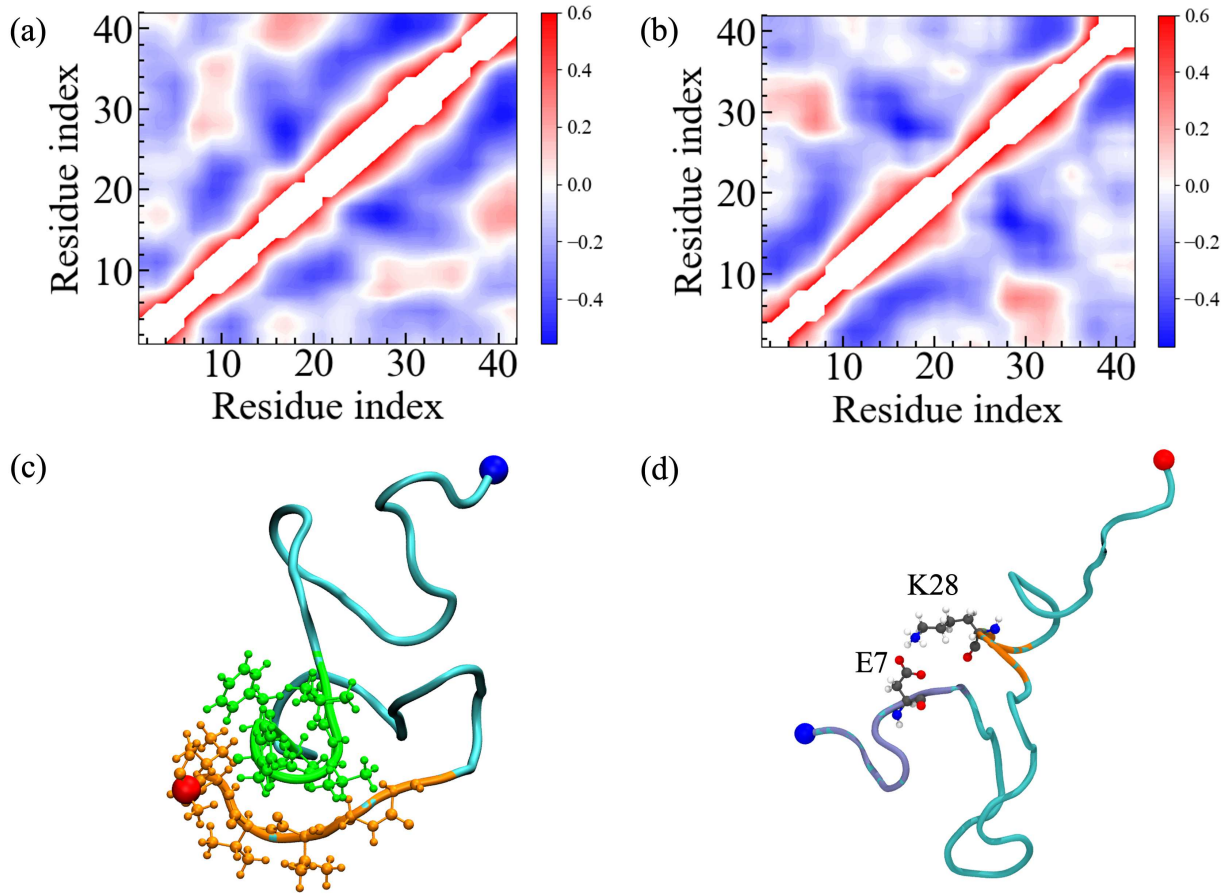}
		\caption{Two-dimensional dynamic cross-correlation map extracted from (a) wild type 
		and (b) mutant A$\beta_{1-42}$ ResET simulations. A representative snapshot obtained from 
		the wild type simulations is shown in (c), where the central hydrophobic core 
		 L$_{17}$VFFA$_{21}$ and the C-terminal hydrophobic 
		 residues G$_{37}$GVVIA$_{42}$ are drawn  in green and orange color, respectively.
		  A corresponding snapshot from the mutant simulation is shown in (d), 
		  where the disordered N-terminus (residues 1-10) and residues 27-31 are colored 
		  in ice-blue and orange, respectively. N- and C-terminal residues 
		  are represented by blue and red spheres.}
		\label{Fig-Cross}
	\end{figure}

		The lower flexibility of the segment 21-37 in the mutant is not correlated with increased secondary structure. Residues
		take dihedral angle values as in a helix  or a $\beta$ strand with about 10\% in both wild type and mutant. However,
		there is a change in the average radius of gyration (RGY, a measure for the volume), which with 10.6(1) \AA\ is larger for the
		mutant than for the wild type where it is 10.5(1) \AA.  Similarly is the average solvent accessible 
		surface area (SASA) of the peptide  in the mutant with
		38.0(1) nm$^2$  less fluctuating than in the wild type (38.0(3) nm$^2$), reflecting the gain in surface area resulting 
		from the more bulky Threonine.  However,  the relation is different for the segment 
		of residues 21-37, where the wild type has a SASA value of 18.2(3) nm$^2$   and the mutant a SASA of 18.1(2) nm$^2$ . 
		The differences for the 
		segment result from polar residues as the solvent accessible surface area of hydrophobic residues
		 is with 4.1(1) nm$^2$  the same for both mutant and wild type.
		Hence, the differences in SASA values for this segment indicate that in the mutant polar residues,  which  are exposed 
		to solvent in the wild type,  form contacts with other residues.
		In order to understand the differences between mutant and wild type in more detail, we have also analyzed the contacts and 
		cross-correlations between residues, focusing again on the final 50 ns of the trajectories for both systems. The resulting maps
		for both systems are shown in  Figure~\ref{Fig-Cross} a-b, with the coloring describing  
		the degree of correlation between residues.		
		
		Unlike in the wild type are in the A2T mutant  the disordered N-terminus (residues 1-9)  and residues 27-33   correlated. 
		This correlation results    from electrostatic interactions, for instance between the
		 NH3+ group of residue K28  (a Lysin)  with negatively-charged COO- group of residue  E7 (a Glutamic acid) 
		   seen in the snapshot shown in Figure   \ref{Fig-Cross} d.  
		 Hence, the replacement of the small 
		 hydrophobic Alanine by a bulky polar Threonine allows for the above  electrostatic interactions	in the mutant that do not exist
		 in the wild type, and whose importance  for inhibiting amyloid formation in the A2T mutant  
		has been already noticed earlier  in Ref.\cite{das_2015}. These interactions likely  stabilize not only the segment 27-33,
		but are responsible for the lower RMSF seen for residues 21-37.
		    The   interactions between N-terminus and residues 27-33
		compete now in the A2T with hydrophobic  interactions between the   segment formed by residues 13-21, which include 
		the central hydrophobic core (L$_{17}$VFFA$_{21}$), and the mostly hydrophobic C-terminus (residues 37-42),  
		 see the corresponding snapshot in Figure \ref{Fig-Cross} c. As a result the two segments are correlated in the wild type 
		 but not in the mutants. These interactions  between the peptide's two main  hydrophobic domains  are 	thought 
		 to be crucial   for the self-assembly 
                  of A$\beta$-fibrils \cite{zhang_2010, bernstein_2005},  but are now missing in the A2T mutant, reducing the risk for aggregation.

	\newpage	
	
  	\section{Conclusions}
  	We have described a  replica-exchange-based multiscale simulation method, Resolution-Exchange with Tunneling (ResET),
	designed for simulation of protein-folding and aggregation. Our approach combines 
  	 the faster sampling in coarse-grained simulations with the potentially higher accuracy 
	of all-atom simulations. It avoids the problem of low acceptance rates plaguing similar approaches 
	and requires only few replica. After testing the accuracy and efficiency of our approach for the  
	small Trp-cage protein by comparing our approach  with long-scale (5 $\mu$s) regular molecular dynamic simulations, 
	we use our new method to compare to compare the ensemble of A$\beta_{1-42}$ wild type peptides, 
	implicated in Alzheimer's disease,  with that of  A2T mutants which seems to protect
	against Alzheimer's disease.	Our ResET simulations indicate  that the replacement of a small Alanine (A)
	by a bulky Threonine (T) as  residue 2 alters the pathway for amyloid formation by introducing steric constraints 
	on the mostly polar N-terminal residues that encourage electrostatic interactions with residues 27-33. These 
	interactions  reduce the flexibility of the extended segment 21-37, therefore contributing to the overall larger volume,
	more exposed surface and resulting higher solubility of the mutant. At the same time do this interactions also interfere with the hydrophobic 
	interactions between the central hydrophobic core (L$_{17}$VFFA$_{21}$), and the mostly hydrophobic C-terminus 
	(residues 37-42), known to be crucial for the self-assembly of A$\beta$-fibrils, decreasing therefore 	
	 the chance of formation  of A$\beta$-amyloids. Further contributing to   this mechanism that may explain why
	the A2T mutant seems to protect the carrier against Alzheimer's disease, could be  the larger exposed
	hydrophobic surface area that in connection with increase solubility may trigger faster degradation of the mutant. 	
	We plan to test this hypothesis by comparing the A2T mutant with suitable double mutants  that interfere with this 
	mechanism.
				
\begin{acknowledgments}
The simulations in this work were done
using the SCHOONER cluster of the University of Oklahoma, XSEDE resources
allocated under grant MCB160005 (National Science Foundation), and TACC resources 
allocated under grant under grant MCB20016  (National Science Foundation). We acknowledge
financial support from the National Institutes of Health under grant GM120634 and GM120578. 
FY acknowledges support by the Scientific and Technological Research Council 
 of TURKEY (TUBITAK) under the BIDEB programs. We are grateful to Dr. Nathan Bernhardt
 for contributions   at an early stage of this project.  F.Y. also thanks the Department of Chemistry 
 and Biochemistry for kind hospitality during his sabbatical stay at the University of Oklahoma.
 \end{acknowledgments}
 
 \bibliographystyle{apsrev4-2} 	
  \bibliography{ResET}

\end{document}